\documentstyle[12pt]{article}
\newcommand{\Z}{{\sf Z \!\!\! Z}}

\newcommand{\DPsi}{{\cal D}\Psi}
\newcommand{\DPsibar}{{\cal D}\overline{\Psi}}
\newcommand{\Psibar}{\overline{\Psi}}
\newcommand{\Deta}{{\cal D}\eta}
\newcommand{\Detabar}{{\cal D}\overline{\eta}}
\newcommand{\muhat}{\hat{\mu}}
\newcommand{\etabar}{\overline{\eta}}
\makeatletter
\@addtoreset{equation}{section}
\makeatother
\title{Fixed Point Actions for Wilson Fermions}
\author{U.-J. Wiese \\[2em]
H\"ochstleistungsrechenzentrum (HLRZ), 5170 J\"ulich, Germany}
\begin{document}

\maketitle

\,

\,

\,

\begin{abstract} \normalsize

Iterating renormalization group transformations for lattice fermions the
Wilson action is driven to fixed points of the renormalization group. A
line of fixed points is found and the fixed point actions are computed
analytically. They are local and may be used to improve scaling in lattice
QCD. The action at the line's endpoint is chirally invariant and still
has no fermion doubling. The Nielsen-Ninomiya theorem is evaded because
in this case the fixed point action is nonlocal. The use of this
action for a construction of lattice chiral fermions is discussed.

\end{abstract}

\pagebreak

The universal continuum behavior of a lattice field theory is determined by the
properties of the corresponding fixed point of the renormalization group
\cite{Wil73}. Free field theory is one of the few cases where renormalization
group transformations can be carried out exactly and the fixed point action can
be investigated in detail. For a free lattice scalar field this has been done
by Bell and Wilson \cite{Bel75}. Here a similar study is presented for free
lattice fermions. Recently, Hasenfratz and Niedermayer \cite{Has93} have used
a local fixed point action to improve scaling in asymptotically free lattice
scalar field theory.
It is
natural to apply the same ideas to fermionic systems, e.g. to lattice QCD.
It turns out that the fermionic fixed point action is local as long as the
renormalization group transformation breaks chiral symmetry.
The locality properties of actions obtained from renormalization
group transformations of lattice fermions have also been studied
by Balaban, O'Carroll and Schor \cite{Bal89}.

Lattice fermions suffer from the well-known doubling problem. Wilson
has removed the doubler fermions by adding an irrelevant chiral symmetry
breaking term to the action \cite{Wil74}. This gives the doublers a mass of the
order of the cut-off and removes them from the physical spectrum. Then,
in the cut-off field theory chiral symmetry is explicitly broken. In fact,
the Nielsen-Ninomiya theorem excludes a chirally invariant solution of the
doubling problem assuming hermiticity, locality and translation invariance of
the fermion action \cite{Nie81}. This prevents the lattice formulation of
chiral
gauge theories (like the Standard Model) because then chiral breaking terms
destroy gauge invariance. For free chiral fermions doubling can be avoided
in various ways. For example, one may use SLAC fermions \cite{Dre76} which
have a nonlocal action. However, when interactions are included the nonlocality
causes severe problems already in perturbation theory \cite{Kar78}. When
a chirally symmetric renormalization group transformation
is used the resulting fixed point action is chirally invariant, and it can be
used to describe free chiral fermions. This
does not (yet) solve the problems of lattice chiral gauge theories,
but it suggests an interesting line of thought.

The partition function $Z = \int \DPsibar \DPsi \exp(- S[\Psibar,\Psi])$
for free Wilson fermions is determined by the action
\begin{eqnarray}
S[\Psibar,\Psi]&=&\frac{1}{2} \sum_{x,\mu}
(\Psibar_x \gamma_\mu \Psi_{x+\muhat} - \Psibar_{x+\muhat} \gamma_\mu \Psi_x) +
\sum_x m \Psibar_x \Psi_x \nonumber \\
&+&\frac{r}{2} \sum_{x,\mu}
(2 \Psibar_{x} \Psi_x - \Psibar_x \Psi_{x+\muhat} - \Psibar_{x+\muhat} \Psi_x).
\label{action}
\end{eqnarray}
Here $\Psibar_x$ and $\Psi_x$ are independent Grassmann valued spinors living
on the sites $x$ of a $d$-dimensional cubic lattice $\Lambda$, $\muhat$ is the
unit vector in $\mu$-direction, and $\gamma_\mu$ are Dirac matrices. The
mass of the physical fermion is given by $m$. The expression proportional to
$r$
is the chiral symmetry breaking Wilson term that gives large masses to the
doubler fermions. In momentum space the corresponding inverse fermion
propagator
is given by
\begin{equation}
\Delta^{-1}(k) = i \sum_\mu \sin k_\mu \, \gamma_\mu + m +
\frac{r}{2} \sum_\mu \left(2 \sin \frac{k_\mu}{2}\right)^2.
\end{equation}
The lattice theory is critical for $m = 0$ where it describes a single massless
Dirac fermion.

A renormalization group transformation maps the theory on the original lattice
$\Lambda$ to a theory on a blocked lattice $\Lambda'$ with doubled lattice
spacing. The points
$x' \in \Lambda'$ of the blocked lattice correspond to cubic blocks of $2^d$
points $x \in \Lambda$ of the original lattice, such that each point $x$
belongs to exactly one block $x'$ (this is denoted by $x \in x'$).
The geometric situation is illustrated in
fig.1.
\begin{figure}
\vspace{7cm}
\caption{The points $x \in \Lambda$ of the original lattice (crosses) together
with the points $x' \in \Lambda'$ of the blocked lattice (circles).}
\end{figure}
On the blocked lattice one defines new fermion fields $\Psibar'$ and
$\Psi'$. The renormalization group step corresponds to the identity
$Z = \int \DPsibar' \DPsi' \exp(- S'[\Psibar',\Psi'])$ with
\begin{eqnarray}
&&\!\!\!\!\!\exp(- S'[\Psibar',\Psi']) = \nonumber \\
&&\!\!\!\!\!\int \DPsibar \DPsi \exp(- a \sum_{x' \in \Lambda'}
(\Psibar'_{x'} - \frac{b}{2^d} \sum_{x \in x'} \Psibar_x)
(\Psi'_{x'} - \frac{b}{2^d} \sum_{x \in x'} \Psi_x))
\exp(- S[\Psibar,\Psi]) = \nonumber \\
&&\!\!\!\!\!\int \DPsibar \DPsi \Detabar \Deta
\exp(\sum_{x' \in \Lambda'}
((\Psibar'_{x'} - \frac{b}{2^d} \sum_{x \in x'} \Psibar_x) \eta_{x'} +
\etabar_{x'}(\Psi'_{x'} - \frac{b}{2^d} \sum_{x \in x'} \Psi_x) +
\frac{1}{a} \etabar_{x'} \eta_{x'}))
\nonumber \\
&&\!\!\!\!\!\times \exp(- S[\Psibar,\Psi]).
\label{step}
\end{eqnarray}
In the last step auxiliary fields $\etabar$, $\eta$ have been introduced.
For $a = \infty$ the renormalization group transformation is chirally
symmetric, because then the chiral breaking $\etabar_{x'} \eta_{x'}$ term
disappears.
The parameter $b$ renormalizes the fermion field. Later it turns out
that $b$ must be set to a special value in order to reach a fixed point of the
renormalization group.
When iterated starting from a point on the critical surface ($m = 0$), the
renormalization group transformation generates a sequence
of actions
\begin{equation}
S[\Psibar,\Psi] \rightarrow S'[\Psibar',\Psi'] \rightarrow
S''[\Psibar'',\Psi''] \rightarrow ... \rightarrow S^*[\Psibar,\Psi],
\end{equation}
that eventually converges to a fixed point action $S^*[\Psibar,\Psi]$.
Of course, also the fixed point is located on the critical surface.
To find the fixed point action let us make the ansatz
\begin{equation}
S[\Psibar,\Psi] = i \sum_{x,y} \rho_\mu(x-y) \Psibar_x \gamma_\mu \Psi_y +
\sum_{x,y} \lambda(x-y) \Psibar_x \Psi_y.
\end{equation}
Charge conjugation invariance requires $\rho_\mu(-z) = - \rho_\mu(z)$ and
$\lambda(-z) = \lambda(z)$. Now we go to momentum space (the Brillouin zone
$B = ]-\pi,\pi]^d$)
\begin{equation}
\rho_\mu(z) = \frac{1}{(2 \pi)^d} \int_B d^dk \,\rho_\mu(k) \exp(i k z),
\,\,\,\, \lambda(z) = \frac{1}{(2 \pi)^d} \int_B d^dk \, \lambda(k) \exp(i k
z).
\end{equation}
In fig.2 the function $\rho_1(k)$ for the Wilson fermion action is displayed
together with
the equipotential lines of $\rho(k)^2$ in the $d = 2$ case.
\begin{figure}
\vspace{20cm}
\caption{(a) The function $\rho_1(k) = \sin k_1$ for the original Wilson
fermion action
over the Brillouin zone $B$, and (b) the equipotential lines of $\rho(k)^2$ in
$B$.}
\end{figure}
After transforming
eq.(\ref{step}) to momentum space
we first perform the Gaussian integration over $\Psibar$, $\Psi$.
Integrating also over $\etabar$, $\eta$ one obtains the action
$S'[\Psibar',\Psi']$. Iterating this procedure $n$ times leads to the action
$S^{(n)}[\Psibar^{(n)},\Psi^{(n)}]$ that is given in terms of
$\rho^{(n)}_\mu(k)$ and $\lambda^{(n)}(k)$. Introducing
\begin{equation}
\alpha_\mu(k) = \frac{\rho_\mu(k)}{\rho(k)^2 + \lambda(k)^2}, \,\,\,\,
\beta(k) = \frac{\lambda(k)}{\rho(k)^2 + \lambda(k)^2}
\end{equation}
and analogously $\alpha^{(n)}_\mu(k)$, $\beta^{(n)}(k)$ one finds
\begin{eqnarray}
&&\!\!\!\!\!\!\!\!
\alpha^{(n)}_\mu(k) = \left(\frac{b^2}{2^d}\right)^n \sum_l
\alpha_\mu(\frac{k + 2 \pi l}{2^n})
\prod_\nu \left(\frac{\sin(k_\nu/2)}
{2^n \sin((k_\nu + 2 \pi l_\nu)/2^{n+1})} \right)^2, \nonumber \\
&&\!\!\!\!\!\!\!\!
\beta^{(n)}(k) = \left(\frac{b^2}{2^d}\right)^n \sum_l
\beta(\frac{k + 2 \pi l}{2^n})
\prod_\nu \left(\frac{\sin(k_\nu/2)}
{2^n \sin((k_\nu + 2 \pi l_\nu)/2^{n+1})} \right)^2 +
\frac{1 - (b^2/2^d)^n}{a(1 - b^2/2^d)}. \nonumber \\ \,
\label{fixed}
\end{eqnarray}
The summation extends over vectors $l$ with integer components
$l_\mu \in \{1,2,...,2^n\}$. Fig.3 shows the situation after two
renormalization group steps for $a = \infty$.
The function $\rho^{(2)}_1(k)$ is depicted
together with the equipotential lines of $\rho^{(2)}(k)^2$.
\begin{figure}
\vspace{20cm}
\caption{(a) The function $\rho^{(2)}_1(k)$ after two renormalization group
steps together with (b) the equipotential lines of $\rho^{(2)}(k)^2$.}
\end{figure}
In the limit $n \rightarrow \infty$ we expect to
approach a fixed point of the renormalization group. Then only the small $k$
(large distance) behavior of the initial functions $\alpha_\mu(k)$ and
$\beta(k)$ is relevant. For the Wilson fermion action one obtains
\begin{equation}
\alpha_\mu(k) \sim \frac{k_\mu}{k^2(1 + m r) + m^2}, \,\,\,\,
\beta(k) \sim \frac{m + k^2 r/2}{k^2(1 + m r) + m^2},
\end{equation}
and at the critical point $m = 0$ one has $\alpha_\mu(k) \sim k_\mu/k^2$
and $\beta(k) \sim r/2$. Inserting this in eq.(\ref{fixed}) one finds
\begin{eqnarray}
&&\alpha^*_\mu(k) = \lim_{n \rightarrow \infty} \left(\frac{b^2}{2^d}\right)^n
\sum_l
\frac{2^n(k_\mu + 2 \pi l_\mu)}{(k + 2 \pi l)^2}
\prod_\nu \left(\frac{\sin(k_\nu/2)}{k_\nu/2 + \pi l_\nu} \right)^2, \nonumber
\\
&&\beta^*(k) = \lim_{n \rightarrow \infty} \left(\frac{b^2}{2^d}\right)^n
\sum_l \frac{r}{2}
\prod_\nu \left(\frac{\sin(k_\nu/2)}{k_\nu/2 + \pi l_\nu} \right)^2 +
\frac{1 - (b^2/2^d)^n}{a(1 - b^2/2^d)}.
\end{eqnarray}
The renormalization group transformations converge to a fixed point only if
\begin{equation}
\left(\frac{b^2}{2^d}\right)^n 2^n = 1 \, \Rightarrow \, b = 2^{(d-1)/2}.
\end{equation}
This is similar to the scalar field case of Bell and Wilson \cite{Bel75},
although $b$ must then be fixed to a different value. Note that $(d-1)/2$
is the canonical dimension of a free fermion field in $d$ dimensions. Hence
$b$ renormalizes the fermion field on the blocked lattice by the appropriate
amount. The fixed point action is determined by
\begin{equation}
\alpha^*_\mu(k) = \sum_{l \in \Z^d}
\frac{k_\mu + 2 \pi l_\mu}{(k + 2 \pi l)^2}
\prod_\nu \left(\frac{\sin(k_\nu/2)}{k_\nu/2 + \pi l_\nu} \right)^2, \,\,\,\,
\beta^*(k) = \frac{2}{a}.
\label{fix}
\end{equation}
As in the scalar field case \cite{Bel75} there is a line of fixed points
parametrized by $a$.
For finite $a$ both the renormalization group transformation and
the fixed point action break chiral symmetry. As a consequence, the doubler
fermions are removed by a Wilson term $\lambda^*(k) \neq 0$ and the fixed
point action is local (exponentially suppressed at large distances).
At the line's endpoint $a = \infty$ one obtains
$\rho^*_\mu(k) = \alpha^*_\mu(k)/\alpha^*(k)^2$
and $\lambda^*(k) = 0$, i.e. the Wilson term disappears at the fixed point.
Strictly speaking, $\lambda^*(k)$ is singular at the locations of the former
doubler fermions. Putting it to zero also at these isolated points still
corresponds to a fixed point of the renormalization group.
When $\lambda^*(k)$ vanishes the fixed point action
is chirally invariant. Hence, based on the Nielsen-Ninomiya theorem one might
suspect that the doubler fermions reappear at the fixed point. Fortunately,
this
is not the case because the fixed point action is nonlocal and the theorem does
not apply. The function $\rho^*_1(k)$ and the
equipotential lines of $\rho^*(k)^2$ are shown in fig.4 for $a = \infty$.
\begin{figure}
\vspace{20cm}
\caption{(a) The function $\rho^*_1(k)$ for the fixed point action
together with (b) the equipotential lines of $\rho^*(k)^2$.}
\end{figure}
At the fixed point there is only one zero of $\rho^*(k)^2$ and hence only one
physical fermion. In fact, due to the nonlocality of the fixed point action the
inverse fermion propagator
$\Delta^{-1}(k) = i \sum_\mu \rho^*_\mu(k) \gamma_\mu$ diverges at the
locations of the former doubler fermion zeros.
Consequently, the topological argument of
Nielsen and Ninomiya does not apply because it assumes regularity of the
inverse
propagator. Fig.4 also shows that for the fixed point action rotation
invariance
is restored to a larger extent than for the original Wilson action that was
depicted in fig.2 (the region of
almost circular equipotential lines is much bigger). Unlike chiral symmetry,
rotation invariance is not completely restored at the fixed point. Of course,
the
fixed point action has infinite correlation length, such that the violation of
rotation invariance at the scale of the lattice spacing does not affect
physical
quantities in the continuum limit. The propagator of fixed point
fermions is $\Delta(k) = - i \sum_\mu \alpha^*_\mu(k) \gamma_\mu$. The function
$\alpha^*_1(k)$ shown in fig.5 has a single pole and it is
periodic and continuous over the Brillouin zone.
\begin{figure}
\vspace{10cm}
\caption{The function $\alpha^*_1(k)$ over the Brillouin zone.}
\end{figure}
In fig.6a the function $\rho^*_1(z)$ is displayed in coordinate space for
$a = \infty$.
\begin{figure}
\vspace{20cm}
\caption{(a) The function $i \rho^*_1(z)$ for the fixed point action
in coordinate space at $a = \infty$ and (b) $10 i \rho^*_1(z)$ at
$a = 4$.}
\end{figure}
It decays rather slowly indicating the nonlocality of the fixed point action.
One may optimize the renormalization group transformation by choosing $a$
such that the fixed point action is as local as possible. For $d = 2$ the
optimal value is close to $a = 4$. The corresponding function $\rho^*_1(z)$ is
depicted in fig.6b. Such an optimized action may be used in the spirit of
Hasenfratz and
Niedermayer \cite{Has93} to improve the scaling behavior of asymptotically free
fermionic lattice theories like e.g. QCD or the Gross-Neveu model.

Having obtained the fixed point actions explicitly it is also possible to
determine the values of critical exponents. This requires to introduce small
perturbations around a fixed point
\begin{equation}
S[\Psibar,\Psi] = S^*[\Psibar,\Psi] + \delta S[\Psibar,\Psi],
\end{equation}
and to investigate their behavior under renormalization group transformations.
In particular, we are interested in eigenfunctionals $\delta S[\Psibar,\Psi]$
that reproduce themselves under renormalization such that
\begin{equation}
S'[\Psibar',\Psi'] = S^*[\Psibar',\Psi'] + \delta S'[\Psibar',\Psi'] =
S^*[\Psibar',\Psi'] + \gamma \, \delta S[\Psibar',\Psi'] .
\end{equation}
Relevant perturbations have
eigenvalues $\gamma > 1$. They get amplified under
renormalization group transformations. In the present case one expects one
relevant direction associated with the fermion mass operator
$m\Psibar_x\Psi_x$.
The corresponding critical exponent $\nu$ describes the divergence of the
correlation length $\xi$ as one approaches the critical point $m = 0$, namely
$1/\xi \propto m^\nu$.
For free fermions one has $\sinh(1/\xi) = m$ such that $\nu = 1$.
Alternatively,
the value of $\nu$ is related to the relevant eigenvalue $\gamma$. Because
the renormalization group transformation changes the scale by a factor of 2 one
has
$\gamma^\nu = 2$.
Using $\delta\lambda(k) = m$ one obtains $\beta(k) = 2/a + m/k^2$ for small
$m$.
Iterating this according to eq.(\ref{fixed}) one finds
\begin{equation}
\beta(k) = \frac{2}{a} + \sum_{l \in \Z^d} \frac{m}{(k + 2 \pi l)^2}
\prod_\nu \left(\frac{\sin(k_\nu/2)}{k_\nu/2 + \pi l_\nu} \right)^2 =
\beta^*(k) + \delta\beta(k).
\end{equation}
Under a renormalization group step one finds
$\delta\beta'(k) = (b^2/2^d) 4 \, \delta\beta(k) = 2 \, \delta\beta(k)$.
For small $m$ the contribution to the action is given by
$\delta\lambda(k) = \delta\beta(k)/(\alpha^*(k)^2 + \beta^*(k)^2)$.
Hence, under a renormalization group
step $\delta\lambda'(k) = 2 \, \delta\lambda(k)$ at least for small $m$.
The corresponding eigenvalue is $\gamma = 2$ and indeed $\nu = 1$.

The present investigation shows how free Wilson fermions solve their
doubling problem at a fixed point of the renormalization group. When a
chirally symmetric renormalization group transformation is used the
fixed point action is also chirally invariant. Since it still describes a
single physical
fermion the Nielsen-Ninomiya theorem must be evaded. This is indeed the case
because the fixed point action is nonlocal. In practical applications a
nonlocal
action may cause severe problems because it slows down a numerical simulation.
In principle, however, the nonlocality of the chirally invariant fixed point
action is
acceptable in the framework of local quantum field
theory, because it arises naturally due to the integration over the high
momentum modes of the field. This is in contrast to SLAC fermions
\cite{Dre76} whose nonlocality is put by hand.
Still, the nonlocality of the fixed point action prevents the explicit
construction of a positive transfer matrix. This is no problem here because the
transfer matrix can be constructed for the original Wilson fermion action
\cite{Lue77}, and the spectrum of low energy states is invariant under the
renormalization group. The fixed point action can be used to put free
chiral fermions on the lattice, simply by keeping, for example, the left-handed
spinors only. Of course, the question arises if the
nonlocality of the action is still acceptable when interactions are
included. In particular, the resulting continuum theory should be local and
Lorentz-invariant. The existence of a well-behaved continuum limit is related
to Reisz's lattice power counting theorem \cite{Rei88}. The theorem assumes
lattice propagators that have a single pole, are periodic over the Brillouin
zone, and are differentiable often enough. For example, SLAC fermions are
excluded because their propagator is discontinuous at the Brillouin zone
boundary. Fixed point fermions, on the other hand, have a propagator
$\Delta(k) = - i \sum_\mu \alpha^*_\mu(k) \gamma_\mu$ that is periodic and
differentiable infinitely many times (as one sees in fig.5). Still, in
its present form the theorem does not apply, because the inverse propagator
$\Delta^{-1}(k) = i \sum_\mu \rho^*_\mu(k) \gamma_\mu$ has isolated poles at
the boundary of the Brillouin zone. Since the poles correspond to zeros of the
propagator they should not cause trouble. It remains to be seen if Reisz's
theorem can be slightly generalized such that it applies to fixed point
fermions.
Of course, the properties of the free fermion propagator are not sufficient
to decide about the interacting theory. Also the vertex functions must be
well-behaved. In this respect, an interesting lesson may be learnt from the
chiral fermion proposal of ref.\cite{Reb87} and how it failed \cite{Bod87}.
To decide if interacting chiral fermions can be constructed with a fixed point
action, one should switch on gauge couplings or four-fermion interactions.
This is presently under investigation in the context of the Gross-Neveu model.

An alternative formulation of lattice fermions uses staggered fermion fields
\cite{Sus77}.
This has the advantage that the cut-off theory has a remnant chiral symmetry.
Blocking transformations consistent with the symmetries of staggered fermions
have been discussed in ref.\cite{Kal92}. An investigation of
the fixed point action for staggered fermions is in progress. A very
interesting
application of fixed point actions has been suggested by Hasenfratz and
Niedermayer to improve scaling in asymptotically free field theories. In QCD,
for example, one could use an optimized local fixed point action with finite
$a$
which then
necessarily breaks chiral symmetry. Of course, this also requires the inclusion
of $SU(3)$ gauge fields.

It is a pleasure to thank W. Bietenholz, M. G\"ockeler, P. Hasenfratz,
J. Jers\'{a}k, F. Niedermayer and T. Reisz for interesting discussions and
C. Rebbi for a helpful remark.

\end{document}